\documentclass[10pt]{iopart}
\usepackage{epsfig}
\usepackage{bm}

\begin{document}

\title[Rapidity dependent momentum anisotropy at RHIC]{Rapidity dependent 
momentum anisotropy at RHIC}

\author{Ulrich Heinz\dag\ and Peter F Kolb\ddag  
}

\address{\dag\ 
Department of Physics, The Ohio State University, % 174 West 18th Avenue, 
Columbus, OH 43210, USA}
\address{\ddag\ 
Physik Department, %James-Franck Strasse, 
TU M\"unchen, D-85747 Garching, Germany}

\begin{abstract}
In Au+Au collisions at RHIC, elliptic flow was found to rapidly decrease 
as a function of rapidity. We argue that the origin of this phenomenon is 
incomplete thermalization of the initial fireball outside the midrapidity
region and show that it can be quantitatively related to the analogous
effect at midrapidity in peripheral collisions and in collisions at 
lower beam energies.
\end{abstract}



%\maketitle

%%%%%%%%%%%%%%%%%%%%%%%%%%%%%%%%%%%%%%%%%%%%%%%%%%%%%%%%%%%%%%%%%%%%%%%%%%
\section{Introduction and overview}
%%%%%%%%%%%%%%%%%%%%%%%%%%%%%%%%%%%%%%%%%%%%%%%%%%%%%%%%%%%%%%%%%%%%%%%%%%

Semicentral Au+Au collisions at RHIC exhibit very strong elliptic flow
$v_2$ which exhausts the prediction from ideal fluid dynamics (for a 
review see \cite{KH03}). This has been interpreted as strong evidence 
for early thermalization of the collision fireball, on a timescale of 
${\,<\,}1$\,fm/$c$ and at energy densities of $\sim 10-20$ times the
critical value for quark-gluon plasma formation \cite{HK02}. However, 
in more peripheral Au+Au collisions at RHIC the measured elliptic flow 
remains increasingly below the hydrodynamic value, and in lower energy
Pb+Pb and Au+Au collisions at the SPS and AGS it does not reach the 
hydrodynamic limit even in central collisions \cite{STARv2PRC,NA49v2PRC}.
The STAR and NA49 Collaborations have suggested (see Fig.\,22 in 
\cite{STARv2PRC} and Fig.\,25 in  \cite{NA49v2PRC}) that the discrepancy
between the hydrodynamically predicted and measured elliptic flow at 
midrapidity scales with $(1/S)\,dN_{\rm ch}/dy$ where $dN_{\rm ch}/dy$ 
is the measured charged multiplicity density at midrapidity and $S$ is 
the initial transverse overlap area between the colliding nuclei. When 
multiplied with the average transverse mass $\langle m_\perp\rangle$ 
per hadron and divided by $\tau_0$, this scaling variable is just the 
Bjorken energy density at proper time $\tau_0$ \cite{Bj}. This suggests 
that the degree of thermalization reached early in the collision (as 
measured by the fraction $v_2^{\rm meas}/v_2^{\rm hydro}$ of the 
hydrodynamically predicted elliptic flow achieved in the experiment) 
is controlled by the initial energy or particle density. This makes 
sense since the collision rate is proportional to the density of 
scatterers.

The PHOBOS Collaboration first observed \cite{PHOBOS02v2eta} that $v_2$ 
shows a strong pseu\-do\-rapidity dependence, with a shape that seems to 
roughly follow the charged particle pseudorapidity distribution. This 
has now been confirmed by STAR \cite{STARv2eta}. BRAHMS and PHOBOS 
also reported that the pseudorapidity distributions of both $N_{\rm ch}$ 
\cite{BRAHMSdNdeta,PHOBOSdNdeta} and $v_2$ \cite{Tonjes} have very 
similar shapes for all collision centralities. The similarity in shape 
between the rapidity distributions of $v_2$ and $N_{\rm ch}$ becomes
even stronger if the Jacobian between rapidity $y$ and pseudorapidity
$\eta$ (which affects $v_2(\eta)$ and $dN_{\rm ch}/d\eta$ mostly near
midrapidity and with opposite signs \cite{Kolb01}) is taken out.    

Hydrodynamic models can reproduce the shape of $dN_{\rm ch}/d\eta$ but 
not that of $v_2(\eta)$ \cite{Hirano01}. Instead of falling off with
increasing rapidity, the hydrodynamically predicted $v_2(\eta)$ actually
first increases with $|\eta|$ before eventually dropping steeply near 
the projectile and beam rapidities \cite{Hirano01,Hirano02}. We find 
two reasons for this behaviour: (i) The initial 
transverse spatial deformation $\epsilon_x$ of the nuclear overlap 
region, which causes the anisotropic pressure gradients driving the 
buildup of elliptic flow, slightly increases with space-time rapidity 
$|\eta_s|{\,=\,}\frac{1}{2}\ln\frac{t+z}{t-z}$, favoring larger $v_2$ at 
$\eta{\,\ne\,}0$. (ii) The initial transversally averaged energy 
density $e(\eta_s;\tau_0) = \langle e(\bm{r},\eta_s;\tau_0)\rangle$ 
decreases with $|\eta_s|$. This reduces the softening effects on the flow
buildup from the quark-hadron phase transition and again results in larger
$v_2$. The same effect leads to larger $v_2$ at midrapidity when one 
reduces the beam energy from RHIC to SPS and AGS energies \cite{KSH00}. 
$v_2$ falls to zero only near the projectile and target rapidities since 
there the initial energy density becomes so small that no time is left 
for flow buildup before freeze-out.

Hirano interpreted the observation that for $\eta{\,\ne\,}0$ the measured 
$v_2$ in Au+Au at RHIC is significantly below the hydrodynamically 
predicted one as evidence that local thermal equilibrium in the initial state 
of the collision is only achieved at midrapidity \cite{Hirano01}. We here
support this conclusion by showing that the ratio $v_2^{\rm meas}/
 v_2^{\rm hydro}$ scales with the initial transversally averaged energy 
density as a function of space-time resp. pseudo-rapidity in the same 
way as previously observed by STAR and NA49 as a function of centrality 
at midrapidity \cite{STARv2PRC,NA49v2PRC}. Both effects indicate that 
the initial thermalization of the system and the validity of 
hydrodynamics are largely controlled by the initial energy or particle
density achieved in the collision.

%%%%%%%%%%%%%%%%%%%%%%%%%%%%%%%%%%%%%%%%%%%%%%%%%%%%%%%%%%%%%%%%%%%%%%%%%%
\section{Initialization and longitudinal structure of the reaction zone}
%%%%%%%%%%%%%%%%%%%%%%%%%%%%%%%%%%%%%%%%%%%%%%%%%%%%%%%%%%%%%%%%%%%%%%%%%%

\begin{figure}
\begin{center}
 \includegraphics[width=6cm,height=4.68cm]{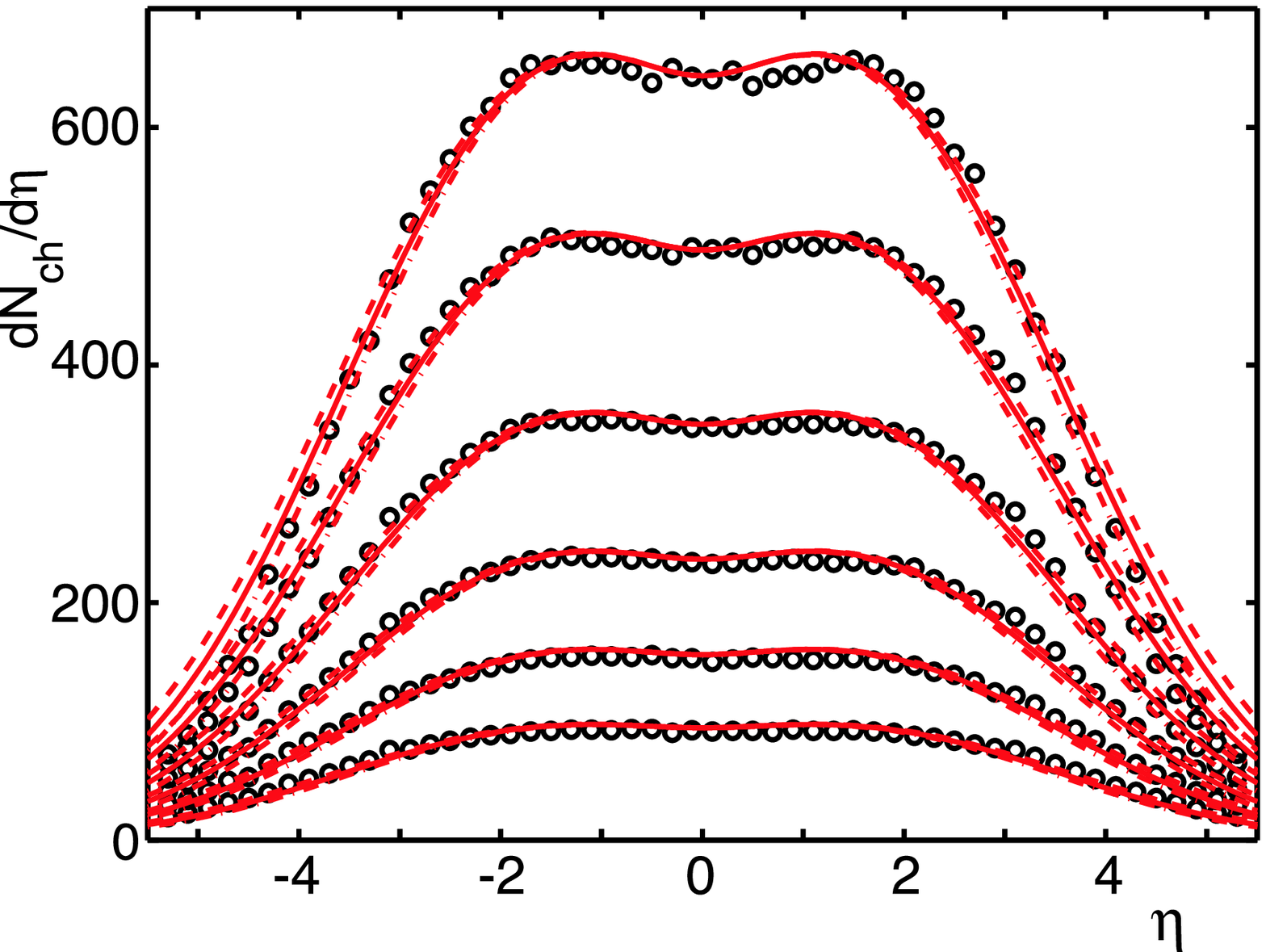} \hspace*{5mm}
 \includegraphics[width=6cm]{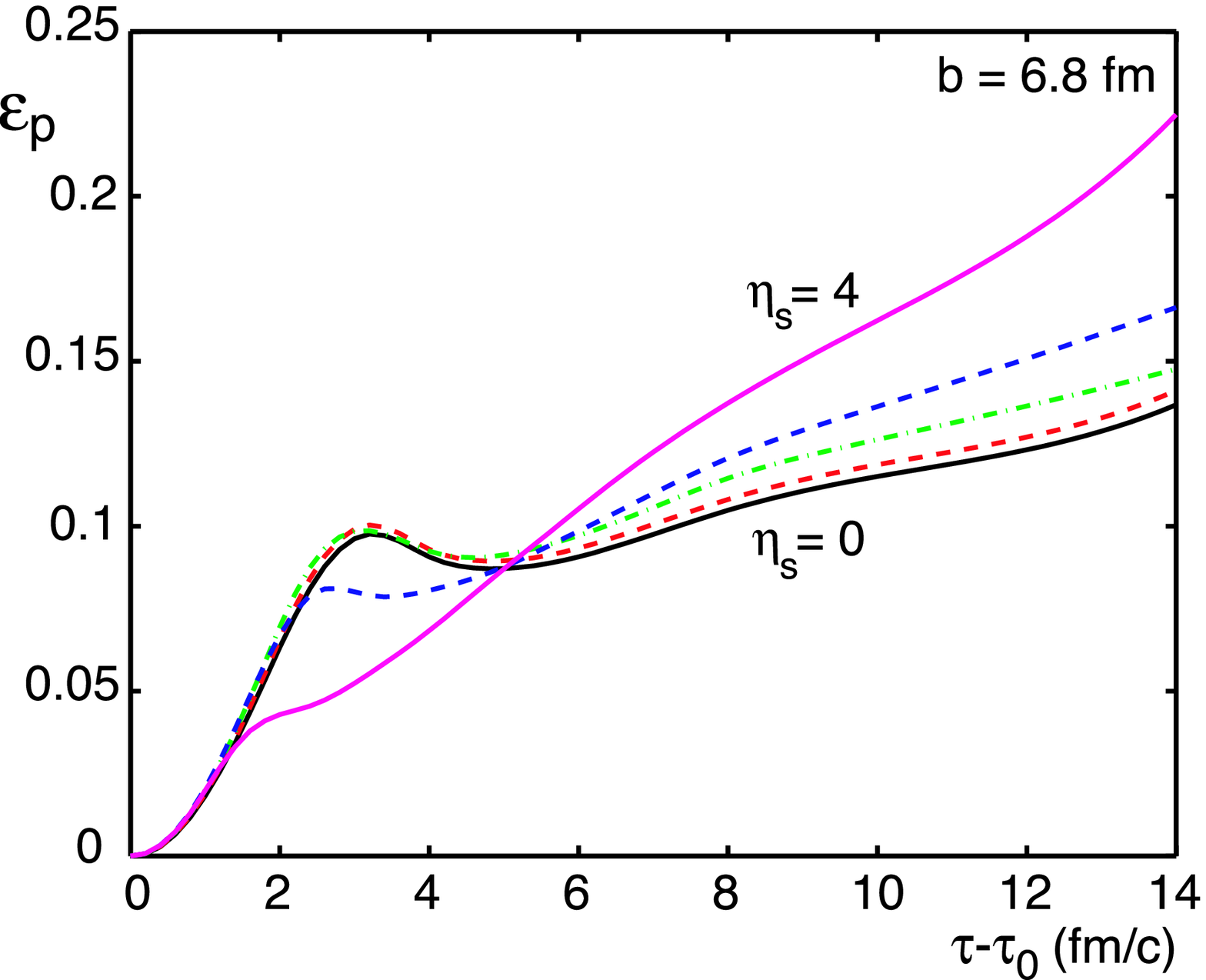}
\end{center}
\caption{\label{F1} {\sl Left:} Pseudorapidity spectra from 200\,$A$\,GeV 
Au+Au collisions with different impact parameters. PHOBOS data
\protect\cite{PHOBOSdNdeta} are compared with a double Gaussian
parametrization (see text). {\sl Right:} Hydrodynamic time evolution of 
the momentum anisotropy $\epsilon_p$ at space-time rapidities
$\eta_s{\,=\,}0,1,2,3,4$.}
\end{figure}

Understanding the rapidity dependence of elliptic flow requires a model
which correlates the initial transverse distribution with longitudinal
position, i.e. with space-time rapidity $\eta_s$. We use Bjorken scaling
$\eta_s{\,=\,}y$ \cite{Bj} during the particle formation stage to relate
initial longitudinal position to initial longitudinal momentum by 
$z{\,=\,}v_{\rm L} t$ on a surface 
$\tau_0{\,=\,}(t^2{-}z^2)^{1/2}{\,=\,}$const. 
The average longitudinal momentum is correlated
with transverse position $\bm{r}$ by overlap geometry coupled with 
momentum conservation \cite{SHR99}: At transverse position $\bm{r}$
two cylinders of matter with thicknesses $t_{1,2}(\bm{r};b)$ 
${=\,}T_{\rm Au}(\bm{r}{\pm}\bm{b}/2)$\footnote[5]{$T_A(\bm{r}) = 
\int_{-\infty}^{\infty}\!\! dz\, \rho_A(\bm{r},z)$ is the 
thickness function of a nucleus of mass $A$, calculated from a 
Woods-Saxon profile $\rho_A(r) = \rho_0/[\exp((r{-}R_A)/\delta)+1]$,
with $R_{\rm Au}{\,=\,}6.37$\,fm and $\delta{\,=\,}0.54$\,fm.} hit each
other, carrying total energy $\gamma_{\rm beam}(t_1{+}t_2)$ and net 
$z$-mo\-men\-tum $\gamma_{\rm beam}v_{\rm beam}(t_1{-}t_2)$. In a collision 
with impact parameter $b$, the produced matter thus has a 
rapidity distribution centered at the $\bm{r}$-dependent avarage 
rapidity \cite{SHR99}   
\begin{equation}
\label{1}
 \fl\qquad
  Y(\bm{r};b) = \frac{1}{2} \ln
  \frac{(t_1(\bm{r};b)+t_2(\bm{r};b)) + v_{\rm beam} 
        (t_1(\bm{r};b)-t_2(\bm{r};b))}
       {(t_1(\bm{r};b)+t_2(\bm{r};b)) - v_{\rm beam} 
        (t_1(\bm{r};b)-t_2(\bm{r};b))} \,.
\end{equation}
In coordinate space the center of mass of the matter produced at transverse
position $\bm{r}$ is located at $\eta_s^{\rm cm}(\bm{r}){\,=\,}Y(\bm{r})$. 
We distribute the initial matter symmetrically around this mean space-time 
rapidity as follows: Noting that 1+1 and 3+1 dimensional hydrodynamical 
calculations \cite{Eskola98,Hirano03} show very little dynamical 
evolution in rapidity direction, we simply identify the shape of the 
initial entropy distribution $dS/d\eta_s$ with that of the measured 
final charged particle rapidity distribution $dN_{\rm ch}/dy$. We 
parametrize the latter by a double Gaussian,
\begin{equation}
\label{2}
  dN_{\rm ch}/dy \propto 
  e^{- \frac{(y-a)^2}{2a^2}} + e^{-\frac{(y+a)^2}{2a^2}},
\end{equation}
with a free width parameter $a$. Combining Eqs.~(\ref{1},\ref{2}) 
and noting that the normalization of the 
rapidity densities at midrapidity scales roughly with the number of 
wounded nucleons, we arrive at the following form for the initial 
entropy distribution:
\begin{equation}
\label{3}
  \fl\quad
  s(\bm{r},\eta_s;\,b) \propto 
  \left(n_1^{\rm WN}(\bm{r}; b) + n_2^{\rm WN}(\bm{r}; b) \right)
  \left(e^{- \frac{(\eta_s - Y(\bm{r}{;}b)-a)^2}{2a^2}} 
      + e^{- \frac{(\eta_s - Y(\bm{r}{;}b)+a)^2}{2a^2}}\right).
\end{equation}
Integration over $\bm{r}$ gives the rapidity distribution 
$dS/d\eta_s\propto dN_{\rm ch}/dy$, which can be transformed to 
pseudorapidity, $dN_{\rm ch}/d\eta$, by multiplying with
the Jacobian $\langle p_\perp\rangle \cosh\eta/(m_\pi^2{+}\langle 
p_\perp\rangle^2 \cosh^2\eta)^{1/2}$, where we treat all charged 
hadrons as pions with $\langle p_\perp\rangle\approx 0.4$\,GeV. A 
single overall normalization constant and the width $a$ of the double
Gaussian (\ref{2}) are fitted to the data. The left panel in 
Fig.~\ref{F1} shows such a fit for 200\,$A$\,GeV Au+Au data
with $a{\,=\,}1.9{\pm}0.05$. Similar fits work well for Au+Au data at 
$\sqrt{s}{\,=\,}130$ and 19.6\,$A$\,GeV, with $a{\,=\,}1.8{\pm}0.05$
and $1.15{\pm}0.05$, respectively. Note that the double Gaussian ansatz
does {\em not} work for d+Au collisions.

%%%%%%%%%%%%%%%%%%%%%%%%%%%%% Figure 1 %%%%%%%%%%%%%%%%%%%%%%%%%%%%%%%%%%%%%
%%%%%%%%%%%%%%%%%%%%%%%%%%%%%%%%%%%%%%%%%%%%%%%%%%%%%%%%%%%%%%%%%%%%%%%%%%%%

%%%%%%%%%%%%%%%%%%%%%%%%%%%%%%%%%%%%%%%%%%%%%%%%%%%%%%%%%%%%%%%%%%%%%%%%%%
\section{Time evolution and anisotropic flow observables}
%%%%%%%%%%%%%%%%%%%%%%%%%%%%%%%%%%%%%%%%%%%%%%%%%%%%%%%%%%%%%%%%%%%%%%%%%%

For each value of $\eta_s$, we evolve these initial conditions with a locally
boost-invariant hydrodynamic code \cite{KSH00} with the equation state 
of \cite{KR03} to simulate the transverse evolution of the reaction zone. 
Here we represent minimum bias Au+Au data by a fixed average impact 
parameter of $b{\,=\,}6.8$\,fm. The evolution is started at 
$\tau_0{\,=\,}0.6$\,fm/$c$ and stopped at decoupling energy density 
$e_{\rm dec}{\,=\,}0.075$\,GeV/fm$^3$, both independent of rapidity.
As noted by Hirano \cite{Hirano03}, the smaller transverse size of the
fireball at forward rapidities should lead to an earlier decoupling
at higher temperature and smaller transverse flow, causing steeper
$p_\perp$ spectra and smaller $p_\perp$-integrated elliptic flow at 
forward rapidities. The same point was made previously by Teaney
for the smaller collision systems formed in peripheral collisions
at midrapidity \cite{Teaney}. This effect is, however, not enough to 
explain the strong rapidity dependence of the measured $v_2$ 
\cite{Hirano03}, and it is not included in our present calculations.

In the right panel of Fig.~\ref{F1} we show the time evolution of the 
momentum ani\-so\-tro\-py $\epsilon_p = \frac{\langle T^{xx} - T^{yy} \rangle}
 {\langle T^{xx}+T^{yy} \rangle}$ at various rapidities 
$\eta_s$.\footnote[4]{Beyond $\tau-\tau_0{\,\simeq\,}11-12$\,fm/$c$ 
our numerical results for $\epsilon_p$ are affected by matter leaving our
finite grid area and tend to be too large. From \cite{KSH00} we know that 
$\epsilon_p$ saturates beyond $\tau-\tau_0{\,\simeq\,}12$\,fm/$c$.} 
As $\eta_s$ increases, the initial energy density at $\eta_s$ decreases, 
and the time evolution of $\epsilon_p$ follows the same pattern as previously 
observed at midrapidity when reducing the collision energy (see Fig.~7 
in \cite{KSH00}). 

\begin{figure}
\begin{center}
 \includegraphics[width=5.5cm,height=4.12cm]{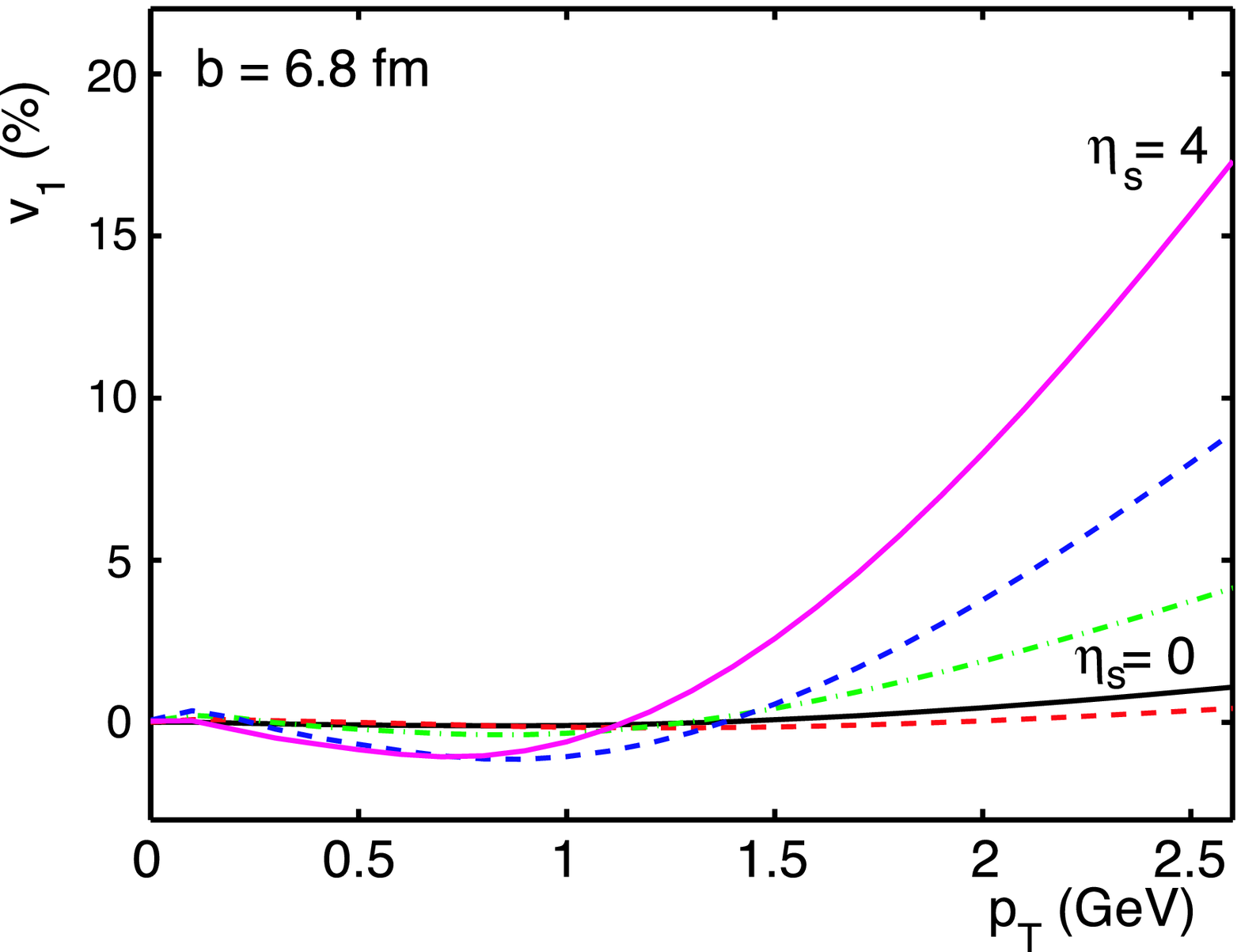}\hspace*{5mm}
 \includegraphics[width=6cm]{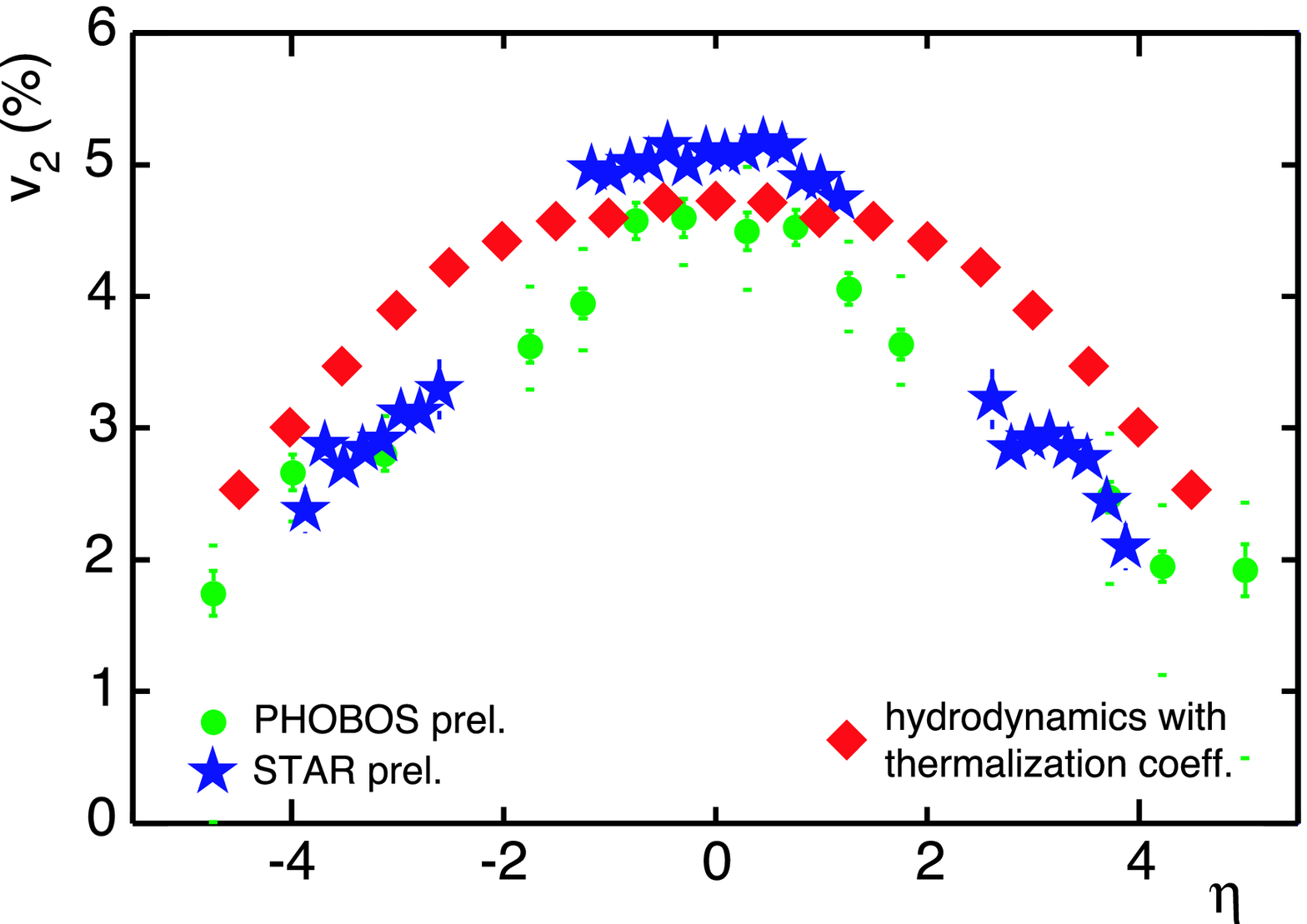}
\end{center}
\caption{\label{F2} {\sl Left:} Differential directed flow $v_1(p_\perp)$
of directly emitted pions (no resonance decays) for 
$\eta_s{\,=\,}y{\,=\,}0,1,2,3,4$. Except for a region of positive 
$v_1$ at $0<p_\perp<0.5$\,GeV and a shift of the rest of the curves by about
0.5\,GeV to larger $p_\perp$, the curves for direct protons look similar.
{\sl Right:} $p_\perp$-integrated elliptic flow $v_2$ for direct pions as
a function of pseudorapidity $\eta$, compared with data for all charged
particles \protect\cite{PHOBOS02v2eta,STARv2eta}. The hydrodynamic $v_2$
values have been corrected with an energy density dependent 
``thermalizations coefficient'' as described in the text. The Jacobian 
for the transformation from $y$ to $\eta$ has been included. 
}
\end{figure}

At forward rapidities the transverse overlap region becomes asymmetric
and is shifted sidewards in the $x$ (or impact parameter) direction.
This turns out to give rise to a non-zero directed flow signal 
$v_1(p_\perp)$ which increases with $|\eta_s|$ (left panel in 
Fig.~\ref{F2}). Of course, since the colliding matter receives no 
overall transverse kick, the $p_\perp$-integrated directed flow is zero.

%%%%%%%%%%%%%%%%%%%%%%%%%%% Figure 2 %%%%%%%%%%%%%%%%%%%%%%%%%%%%%%%%%%%%%%%
%%%%%%%%%%%%%%%%%%%%%%%%%%%%%%%%%%%%%%%%%%%%%%%%%%%%%%%%%%%%%%%%%%%%%%%%%%%%

The hydrodynamically calculated elliptic flow $v_2(\eta)$ has the same 
general shape as previously obtained by Hirano with a fully (3+1)-dimensional 
code. We now correct this hydrodynamic behaviour with a ``thermalization
coefficient'' $F(x)$ which is fitted to midrapidity data in peripheral
and lower-energy collisions \cite{STARv2PRC,NA49v2PRC}. $F$ depends on 
the initial transversally averaged energy density at rapidity 
$y{\,=\,}\eta_s$ through the ratio 
$x(\eta_s){\,=\,}\langle e(\eta_s)\rangle/e_0$ (where 
$e_0{\,=\,}9.5$\,GeV/fm$^3$ is the average initial energy density in 
central Au+Au collisions at 130\,$A$\,GeV). As discussed in the 
Introduction, this scaling variable is, 
up to a multiplicative constant, identical with the variable 
$(1/S)\, dN/dy$ found by STAR and NA49 to control the magnitude of $v_2$ 
at midrapidity \cite{STARv2PRC,NA49v2PRC}. We parametrize the behavior 
shown in Fig.~25 of \cite{NA49v2PRC} with a simple linear function 
$F(x){\,\equiv}$ $\frac{v_2^{\rm meas}}{v_2^{\rm hydro}} = 0.15+0.85\,x$
for $x{\,\leq\,}1$ while $F(x){\,=\,}1$ for $x{\,>\,}1$. ($x{\,=\,}1$ 
corresponds in Fig.~25 of \cite{NA49v2PRC} to 
$(1/S)\, dN_{\rm ch}/dy = 25$\,fm$^{-2}$.)
The corrected $v_2^{\rm meas}(\eta){\,=\,}F(x(\eta))\cdot
v_2^{\rm hydro}(\eta)$ for $b{\,=\,}6.8$\,fm is shown by the full circles 
in the right panel of Fig.~\ref{F2}, together with minimum bias data
from PHOBOS and STAR. Even if our $v_2$ values are still a bit high
at $|\eta|>2$, we see good qualitative agreement with the data. We
conclude that the same incomplete thermalization effects previously
seen at midrapidity in peripheral and lower-energy collisions also
describe qualitatively the rapid decrease of $v_2$ at non-zero rapidity 
in minimum bias collisions at RHIC. Local thermalization seems to 
be driven by the local initial energy density reached in the collision.

%%%%%%%%%%%%%%%%%%%%%%%%%%%%%%%%%%%%%%%%%%%%%%%%%%%%%%%%%%%%%%%%%%%%%%%%%%
\section{Conclusions and further tests}
%%%%%%%%%%%%%%%%%%%%%%%%%%%%%%%%%%%%%%%%%%%%%%%%%%%%%%%%%%%%%%%%%%%%%%%%%%

Our analysis suggests that the critical initial energy density, which 
is required for reaching sufficiently complete local equilibrium during 
the early collision stages to validate a hydrodynamic description, has 
for the first time been reached at midrapidity in almost central Au+Au 
collisions at top RHIC energy. At even higher collision energies
the elliptic flow is therefore expected to follow the hydrodynamic 
predictions over a wider range of rapidities and centralities. However,
there is no need to wait for the LHC to verify this prediction. One way
to test this already at RHIC energies would be to explore fully central
collisions (``zero spectators'') between Uranium nuclei which provide 
slightly larger energy densities than central Au+Au collisions and a 
significant transverse deformation and elliptic flow signal even at 
zero impact parameter, due to the deformation of the projectiles 
\cite{KSH00}. Such full-overlap U+U collisions yield very large, 
transversally deformed fireballs which would also provide a testbed 
for studying more efficiently the predicted nonlinear pathlength 
dependence of jet energy loss (by studying jets as a function of 
their emission angle relative to the reaction plane as extracted from 
$v_2$) and exploring this energy loss out to larger $p_T$ values than 
presently possible with the much smaller deformed fireballs created in 
non-central Au+Au collisions.

\ack
%\medskip

%\noindent{\bf Acknowledgements:} 
This work was supported by US Department of Energy grant DE-FG02-01ER41190
and by DFG and GSI. We thank Tetsu Hirano for valuable discussions.

\section*{References}
\begin{thebibliography}{99}

\bibitem{KH03}
	Kolb P F and Heinz U 2004
	{\it Quark-Gluon Plasma 3} ed R~C~Hwa and X~N~Wang
        (World Scientific, Singapore) ({\it Preprint} nucl-th/0305084)
 %	Hydrodynamic description of ultrarelativistic heavy-ion collisions

\bibitem{HK02}
        Heinz U and Kolb P F 2002
%``Early thermalization at RHIC,''
        {\it Nucl.\ Phys.} A {\bf 702} 269
%%CITATION = HEP-PH 0111075;%%

\bibitem{STARv2PRC}
	Adler C \etal (STAR Collaboration) 2002 
        {\it Phys. Rev.} C {\bf 66} 034904  
% Elliptic flow from two- and four particle correlations in Au+Au at 130 AGeV

\bibitem{NA49v2PRC} 
	Alt C \etal (NA49 Collaboration) 2003 
        {\it Phys. Rev.} C {\bf 68} 034903
% Directed and elliptic flow of charged pions and protons in Pb+Pb at 40 and 158 AGeV 

\bibitem{Bj}
        Bjorken J D 1983 {\it Phys. Rev.} D {\bf 27} 140

\bibitem{PHOBOS02v2eta}
	Back B B \etal (PHOBOS Collaboration) 2002 
        {\it Phys. Rev. Lett.} {\bf 89} 222301
%     ``Pseudorapidity and Centrality Dependence of the Collective Flow of 
%       Charged Particles in Au+Au Collisions at \scm = 130 GeV

\bibitem{STARv2eta}
        Oldenburg M D \etal (STAR Collaboration) 2004 
        {\it Preprint} nucl-ex/0403007

\bibitem{BRAHMSdNdeta}
	Bearden I G \etal (BRAHMS Collaboration) 2002 
        {\it Phys. Rev. Lett.} {\bf 88} 202301

\bibitem{PHOBOSdNdeta}
        Back B B \etal (PHOBOS Collaboration) 2003 
        {\it Phys. Rev. Lett.} {\bf 91} 052303
%	 Significance of the fragmentation region in UltraRHIC

\bibitem{Tonjes}
        Tonjes M B \etal (PHOBOS Collaboration) 2004 
        {\it these proceedings}

\bibitem{Kolb01}
        Kolb P F 2002
%       ``Hydrodynamic flow at RHIC'',
        {\it Heavy Ion Phys.} {\bf 15} 279

\bibitem{Hirano01}
	Hirano T 2001 {\it Phys. Rev.} C {\bf 65} 011901(R)
%	Is early thermalization achieved only near midrapidity 
%	in Au+Au collisions at $\scm=130$~GeV?

\bibitem{Hirano02}
        Hirano T and Tsuda K 2002
 %``Collective flow and two pion correlations from a relativistic  hydrodynamic
%model with early chemical freeze out,''
        {\it Phys. Rev.} C {\bf 66} 054905
%%CITATION = NUCL-TH 0205043;%%

\bibitem{KSH00}
        Kolb P F, Sollfrank J and Heinz U 2000 
        {\it Phys. Rev.} C {\bf 62} 054909
%     ``Anisotropic transverse flow and the quark-hadron phase transition''

\bibitem{SHR99}
	Sollfrank J, Huovinen P and Ruuskanen P V 1999 
        {\it Eur. Phys. J.} C {\bf 6} 525

\bibitem{Eskola98}
        Eskola K J, Kajantie K and Ruuskanen P V 1998
%``Hydrodynamics of nuclear collisions with initial conditions from
% perturbative QCD,''
        {\it Eur.\ Phys.\ J.} C {\bf 1} 627
%%CITATION = NUCL-TH 9705015;%%

\bibitem{Hirano03}
        Hirano T 2004 talk given at RIKEN-BNL Workshop on 
        {\it Collective flow and QGP properties}, BNL, Nov. 17-19, 2003,
        available at http://tonic.physics.sunysb.edu/flow03/

\bibitem{KR03}
        Kolb P F and Rapp R 2003
%``Transverse flow and hadro-chemistry in Au + Au collisions at s(NN)**(1/2) =
%200-GeV,''
        {\it Phys.\ Rev.} C {\bf 67} 044903
%%CITATION = HEP-PH 0210222;%%

\bibitem{Teaney}
        Teaney D, Lauret J and Shuryak E V 2001 
%``A hydrodynamic description of heavy ion collisions at the SPS and RHIC,''
        {\it Preprint} nucl-th/0110037
%%CITATION = NUCL-TH 0110037;%%

\endbib

\end{document}